\documentclass[aps,prl,twocolumn,nofootinbib,superscriptaddress]{revtex4-2}
\pdfoutput=1
\usepackage{graphicx}
\usepackage{dcolumn}   
\usepackage{bm}        
\usepackage{amsmath,amssymb,amsxtra,bm,epsfig}   
\usepackage{float}
\usepackage[dvipsnames]{xcolor}
\usepackage{xspace}
\usepackage{hyperref}
\usepackage{cancel,soul,ulem}
\usepackage{orcidlink}  
\hypersetup{
	colorlinks=true,
	linkcolor=red,
	filecolor=magenta,
	urlcolor=JungleGreen,
	citecolor=RoyalBlue}

\newcommand{\bmt}{\begin{pmatrix}}
	\newcommand{\emt}{\end{pmatrix}}
\newcommand{\ba}{\begin{array}{c}}
	\newcommand{\ea}{\end{array}}
\newcommand{\be}{\begin{equation}}
	\newcommand{\ee}{\end{equation}}
\newcommand{\bea}{\begin{eqnarray}}
	\newcommand{\eea}{\end{eqnarray}}

\newcommand{\bi}{\begin{itemize}}
	\newcommand{\ei}{\end{itemize}}
\newcommand{\baz}{\begin{array}{cc}}
\newcommand{\besub}{\begin{subequations}}
\newcommand{\eesub}{\end{subequations}}

\newcommand{\APPROXIMATE}{\textsc{approximate} }
\newcommand{\ACCURATE}{\textsc{accurate} }

\newcommand{\stk}[1]{\ifmmode\text{\sout{\ensuremath{#1}}}\else\sout{#1}\fi}

\begin{document}

\title{Unveiling a Hidden Epoch: Impact of Mediator Induced Matter Domination in Freeze-in Dark Matter}






\author{Partha Konar \,\orcidlink{0000-0001-8796-1688}}
\email{konar@prl.res.in}
\affiliation{Theoretical Physics Division, Physical Research Laboratory, Ahmedabad, 380009, India}

\author{Sudipta Show \,\orcidlink{0000-0003-0436-6483}}
\email{sudiptas@iitk.ac.in}
\affiliation{Department of Physics, Indian Institute of Technology Kanpur, Kanpur 208016, India}

\begin{abstract}
Freeze-in dark matter has recently garnered significant attention as a promising framework due to its feeble interactions, which are consistent with the null results from dark matter experiments. While previous studies have extensively investigated the production of dark matter through the decay of heavy particles, they overlook the cosmological role of the decaying mediator without justifying this assumption. We emphasize that the mediator can dominate the energy budget of the early universe during its decay, leading to an unavoidable early matter-dominated era. This intrinsic matter-dominated phase influences dark matter production in two key ways: (i) dark matter production occurs in both the early radiation and induced matter-dominated phases; specifically, considerable production occurs in the matter-dominated phase and stops when the mediator decays fully, and (ii) it causes dilution in dark matter abundance due to entropy injection before its saturation. Furthermore, this effect significantly alters the gravitational wave signature associated with the production of freeze-in dark matter through graviton emission during the mediator's decay. Specifically, it enhances the gravitational wave spectrum, making it viable for future high-frequency gravitational wave experiments. 
\end{abstract}


\maketitle

\textbf{\textit{Introduction}---}
Although the ubiquity of dark matter (DM) has been convincingly established through numerous astrophysical and cosmological observations, the nature, production mechanisms, and evolution of DM remain largely elusive, despite extensive efforts over the decades to uncover these puzzles. Specifically, thermal DM in the form of weakly interacting massive particles (WIMPs)~\cite{Arcadi:2017kky, Roszkowski:2017nbc}, once considered to be the most promising and well explored, is under persistent pressure from diverse DM detection experiments~\cite{LZ:2022lsv, PandaX:2024qfu, XENON:2025vwd, Fermi-LAT:2015att, Fermi-LAT:2016uux}.
Subsequently, non-thermal DM, particularly feebly interacting massive particles (FIMPs)~\cite{Hall:2009bx, Bernal:2017kxu}, has gained increasing interest as a viable alternative. In this case, DM never thermalizes with the bath due to its feeble interaction and gets produced gradually from the decay or scattering of bath particles via the so-called freeze-in process.


Notably, freeze-in DM production via decay is a fascinating, well-explored phenomenon, primarily due to its straightforward computation and realization in various scenarios beyond the Standard Model (BSM). Previous studies have largely ignored the consequences of such heavy mediators on the cosmological background, without verifying the validity of this entry-level assumption.
In this letter, we demonstrate that decaying mediator particles with feeble interaction with the DM can dominate the universe's energy budget. Consequently, this often-overlooked but unavoidable mediator domination gives rise to an induced early matter era, thereby significantly altering the production mechanism of DM.

To convince this point, we start with a generic interaction involving a BSM mediator $(X)$, a DM candidate $(\chi)$, and SM particle $(\xi)$, represented by the expression $y X \chi \xi$, which encompasses a broader class of freeze-in DM models.  
The properties of dark matter influence the nature of the BSM and SM particles involved in these interactions. A summary of different well-explored models based on this interaction topology $(y X \xi \chi)$ can be followed from \cite{Konar:2025iuk}.
For this illustration, we further consider models where DM is a scalar particle, the heavy mediator is a vector-like fermion, and the SM particles involved in this interaction are SM fermions.

Many such BSM realizations are being explored at the Large Hadron Collider (LHC) through the direct production of these mediators, thanks to their significant interaction with the SM. A tiny Yukawa interaction with the FIMP candidate can further present distinctive long-lived mediator decay signatures through displaced vertex searches (DV)~\cite{ATLAS:2017tny} and long-lived particle (LLP)~\cite{Curtin:2018mvb} searches.
Nevertheless, such searches can only illuminate a minuscule segment of the vast freeze-in DM parameter space, leaving much of it inaccessible to all current and future experiments including direct and indirect dark matter detection experiments. 

%
%

Fortunately, recent proposals for the new generation of gravitational wave (GW) detection experiments open up a unique window at our disposal, which not only can investigate freeze-in dark matter but also reveal insights into the physics of the early universe. 
The graviton couples with the energy-momentum tensor and interacts with both SM and BSM particles. The emission can occur during the freeze-in production of DM, resulting in a unique GW signature. In reference \cite{Konar:2025iuk}, we demonstrated, for the first time, that the GW signature can be utilized as a probing tool for freeze-in dark matter, followed by additional works \cite{Wang:2025lmf}. 
We will further see that the unavoidable presence of a mediator-dominated matter era significantly enhances the GW signature to the extent that it could be detectable in future experiments.  The remainder of the letter will illustrate how the mediator induced early matter era impacts both the production of DM and the resulting GWs.

\textbf{\textit{Mediator induced matter domination epoch}---} 
Relevant dynamics of the early universe follow coupled Boltzmann equations (BEs), 
\begin{align}
\frac{d\rho_{X}}{dt}+FH\rho_{X}=-\Gamma_{X}\rho_{X},  \; 
\frac{d\rho_R}{dt}+4H\rho_R=+\Gamma_{X}\rho_{X}
	\label{BEs}
\end{align}
where, $\rho_R$ and $\rho_X$ represent the energy density of radiation ($R$) and mediator ($X$) respectively. $\Gamma_X(={y^2m_X}/{(32\pi)})$ and $m_X$ refer to the decay width and mass of $X$ and  mass of $X$ respectively. $H=\sqrt{\rho_R+\rho_X}/(3M_P^2)$ corresponds to the Hubble rate where $M_P(=2.4\times 10^{18}~\text{GeV})$  reduced Planck mass. Note that a re-parameterization of the BEs in terms of scale factor,  $a$, can be achieved using $\frac{d}{da}\equiv\frac{1}{aH}\frac{d}{dt}$. Here, the factor $F$ plays an important role in describing the evolution of energy density of $X$, equal to $4~(3)$ as its energy density falls as $a^{-4}$ ($a^{-3}$) for radiation(matter) domination. Note that $X$ behaves like radiation when it is relativistic $(T>m_X)$ and as matter when non-relativistic $(T<m_X)$.

\begin{figure}[tb!]
	\centering
	\includegraphics[width=\linewidth]{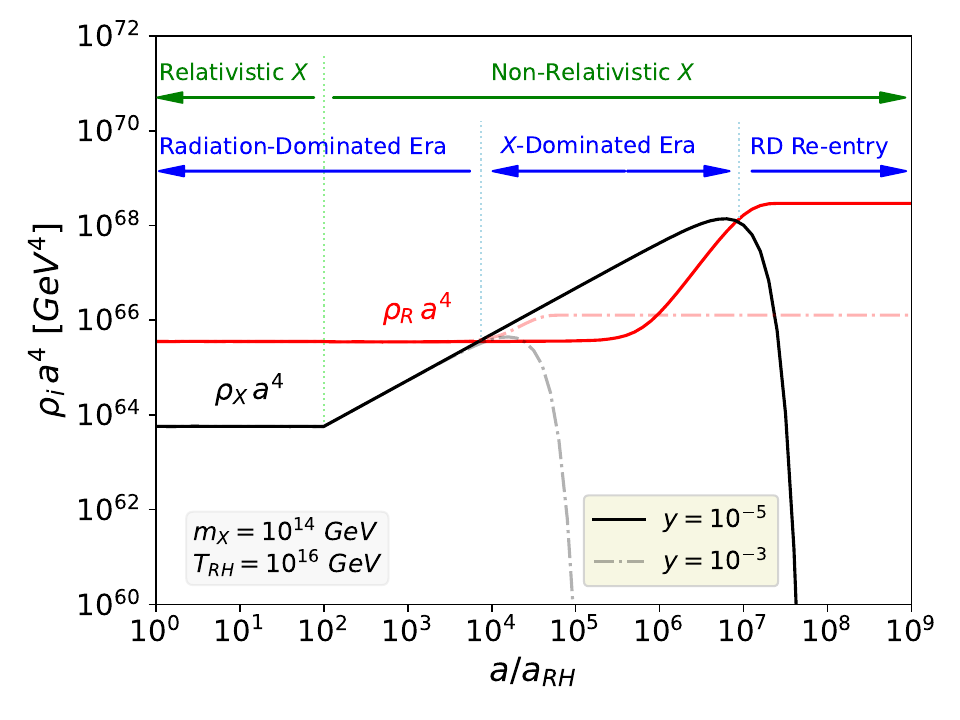}
	\caption{Evolution of the various energy densities as a function of the relative scale factor, $a/a_\text{RH}$.}
\label{Energy_evo}
\end{figure}

In Figure~\ref{Energy_evo}, we present our numerical solution of the coupled Boltzmann equations (BEs), illustrating how a decaying mediator $X$ can trigger an early matter-dominated phase. For this example, we set $y = 10^{-5} $ and $m_X = 10^{14}$ GeV. The solid red and black lines depict the evolution of the energy densities of $R$ and $X $ as functions of the normalized scale factor $a/a_{\text{RH}}$, where $a_{\text{RH}}$ is defined at the reheating temperature, $T_{\text{RH}} (= 10^{16}~\text{GeV})$. Initially, $X$ is relativistic (i.e., $T>m_X$) and behaves like radiation, with its energy density $\rho_X$ scaling as $a^{-4}$. As the universe expands and the temperature drops below the mass of the mediator (i.e., $T<m_X$),  $X$ becomes non-relativistic, causing its energy density to shift to a matter-like scaling, $\rho_X \propto a^{-3} $. During this stage, $X$, which initially had a lower energy density, began to fall at a slower rate than radiation, eventually dominating the energy budget over radiation. As a result, the universe enters a mediator-induced matter-dominated era, which lasts until $X$ has completely decayed. Additionally, we also illustrate the evolution of energy densities for a relatively larger Yukawa coupling $y = 10^{-3}$ as a function of $a/a_{\text{RH}}$  using dot-dashed lines. In this scenario, it becomes evident that $X$ decays just after the matter-dominated era begins for a very short period of time. The universe would not enter into a matter-dominated era for a larger Yukawa coupling.

Interestingly, the criterion for matter domination in terms of Yukawa coupling and mediator mass can be evaluated analytically by requiring that $t_\text{MD}$, demonstrating the time when the matter and radiation energy densities are equal,  is earlier than $t_\text{D}$, dictating the time when the universe becomes radiation dominated again. Here, $t_\text{MD}$ can be expressed as, $	t_\text{MD}=({g_*}/{g_*^X})({M_P}/{2m_X^2})\sqrt{{\pi^2g_*}/{90}}$. Here, $g_*(\simeq106.75)$ and $g_*^X(\simeq1.75)$ refer to the effective number of relativistic degrees of freedom of the SM radiation bath and $X$, respectively. In addition, the mediator domination stops when the energy density equals the radiation energy density for the last (second) time before its full depletion, so the corresponding time $t_\text{D}=\Gamma_X^{-1}$. Now, the analytically estimated criterion for mediator-triggered early matter domination can be written as
\begin{align}
	y^2<0.0465\frac{m_X}{M_P}.
	\label{criterion_MD}
\end{align}
It is obvious from the above condition that there is an upper limit on $y$ for a fixed $m_X$, below which $X$ triggers an early matter domination. However, the smallness of the coupling decides longevity for the course of matter domination since a smaller Yukawa coupling will offer a longer matter-dominated phase.

\textbf{\textit{Dark matter production via freeze-in}---}
In a conventional freeze-in scenario, the DM production occurs gradually through the decay of a heavy particle which remains in thermal equilibrium with bath. The Boltzmann equation governing the evolution of DM number density ($n_\chi(t)$) is given by
\begin{equation}
	\frac{dn_\chi}{dt}+3H n_\chi=\langle\Gamma_X\rangle n_X^{\text{eq}},
	\label{Boltz_nd}
\end{equation}
where, $\langle\Gamma_X\rangle (=\Gamma_X\frac{K_1(m_X/T)}{K_2(m_X/T)})$ represents the thermally averaged decay width, and $n_X^{\text{eq}}(=\frac{g_X m_X^2 T}{2\pi^2}K_2(m_X/T))$ is the equilibrium number density of the decaying particle with $g_X$ being the internal degrees of freedom of $X$. 
After evaluating the final DM number by numerically solving the above equation, the DM relic density can be estimated by using the following expressions
\begin{equation}
	\Omega h^2=2.755\times 10^8 \, m_\chi \, Y_\chi(T\rightarrow 0),
	\label{relic_density}
\end{equation}
where $Y_\chi(n_\chi/s)$ refers to DM abundance with $s$ being the entropy density. Furthermore, one can also evaluate the DM relic density by utilizing the following analytical expression:
\begin{equation}
	\Omega h^2=2.755\times 10^8\frac{M_Pg_X\Gamma_X}{\sqrt{g_*}g_{*s}}\frac{m_\chi}{m_X^2}\frac{3\pi}{2},
	\label{relic_ana}
\end{equation}
Relic density of DM is function of three parameters $y,m_\chi$ and $m_X$, as a consequence, the requirements of the observed DM relic density $(\Omega h^2=0.12)$ will give contours $(\because\Omega h^2\propto y^2 /m_X)$ in Yukawa coupling and mediator plane for any fixed DM masses. Moreover, large Yukawa coupling can bring the DM to thermal equilibrium. 

\begin{figure}[tb!]
	\centering
	\includegraphics[width=\linewidth]{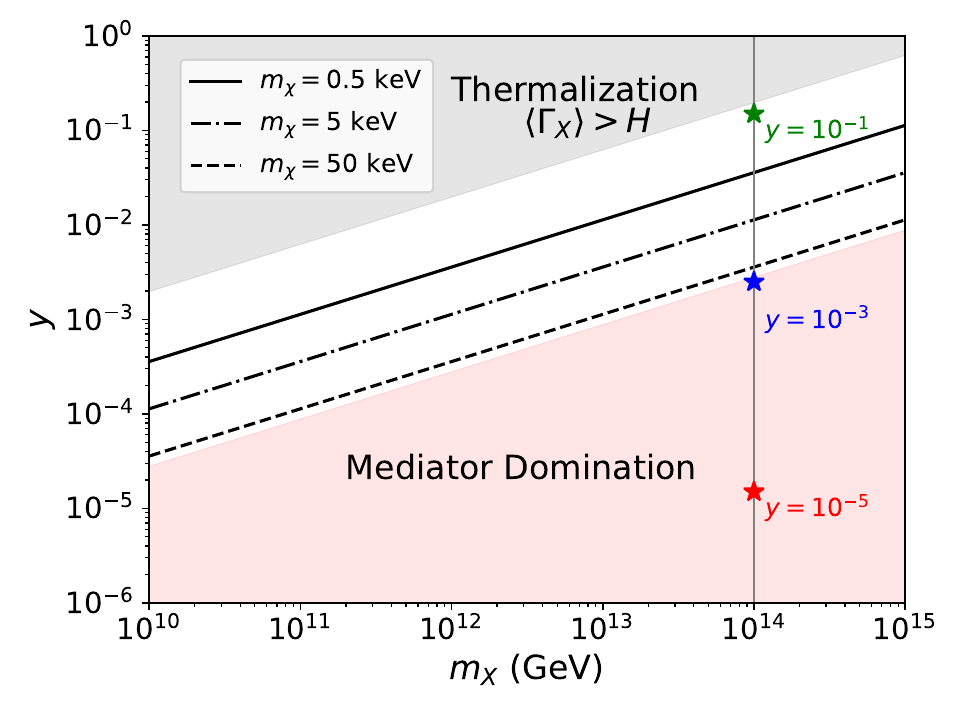}
	\caption{Plot shown the mediator driven matter domination area, thermalization region and the relic density satisfied contours for DM in $y$ and $m_X$ plane.}
	\label{constraint}
\end{figure}

In Figure~\ref{constraint}, we have shown a region in the Yukawa coupling and mediator mass plane for which mediator-driven early matter domination will be realized by the hatched red shaded area using the criterion in Eq.~(\ref {criterion_MD}). The grey shaded hatched region is ruled out since, for these values of $y$ and $m_X$, the interaction rate is larger than the Hubble rate ($\because\langle\Gamma_X\rangle>H$), as a result, DM gets thermalized, thus violating the assumption of freeze-in DM. Additionally, we have shown the relic density satisfied contours by solid, dot-dashed, and dashed lines, representing the DM masses of $0.5$ keV, $5$ keV, and $50$ keV, respectively. Note that, when the mediator mass and Yukawa coupling fall in the red shaded hatched region, the typical estimation of DM relic density using Eq.~(\ref{Boltz_nd}) is not valid because of the induced matter domination. In this case, one has to solve the following Boltzmann equation for DM comoving number density $N_\chi(=n_\chi a^3)$ along with the Eqs.~(\ref{BEs}) to track the evolution of DM.
\begin{align}
	\frac{dN_\chi}{da}=\frac{a^2\Gamma_X}{H(a)}\frac{\rho_X}{E_X},
	\label{Evo_Cov_den}
\end{align}

\begin{figure}[tb!]
	\centering
	\includegraphics[width=\linewidth]{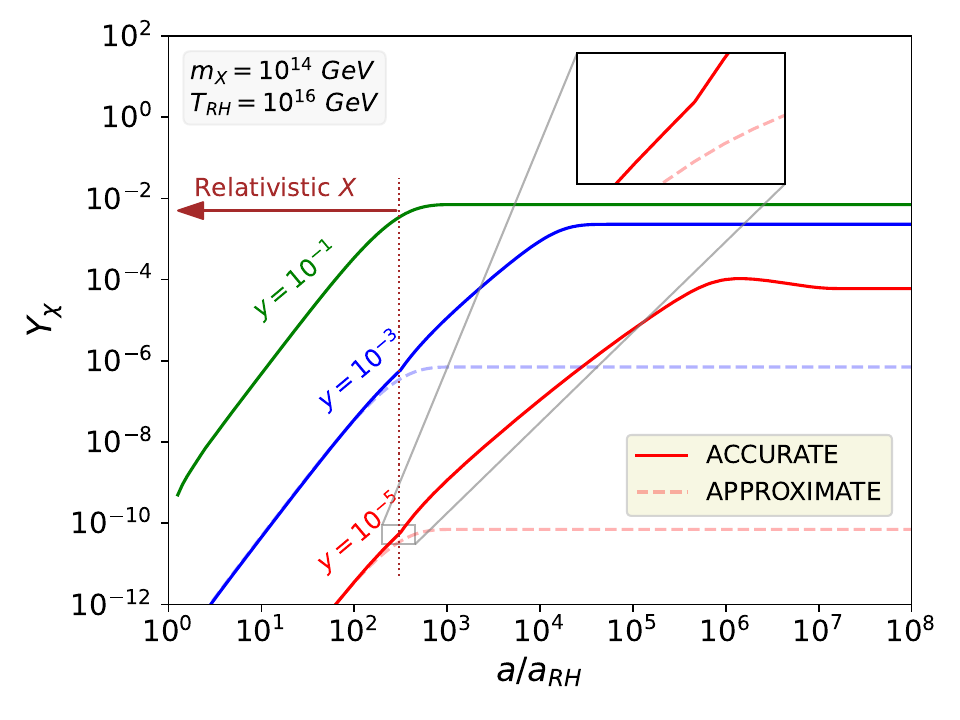}
	\caption{Evolution of DM abundance with $a/a_\text{RH}$.}
	\label{abundance}
\end{figure}
where $E_X(=\sqrt{m_X^2+9T^2})$ corresponds to the energy of $X$. One can obtain the following analytical expression for DM relic density in the case of induced matter domination by using Eq.~{(\ref{Evo_Cov_den})}
\begin{equation}
	\Omega h^2=2.755\times 10^8 m_\chi \times1.6y\times\sqrt{\frac{10^{15}~\text{GeV}}{m_X}}.
	\label{relic_MD}
\end{equation}
For illustration, the impact of mediator domination on dark matter production is demonstrated by numerically solving Eqs. ~(\ref{BEs}) and (\ref{Evo_Cov_den}) simultaneously. From now on, we will refer to results using the usual simplified approach, neglecting the cosmological role of the mediator as ``\APPROXIMATE", and our results incorporating the mediator impacts on cosmological background as ``\ACCURATE". In Figure~\ref{abundance}, the green, blue, and red lines describe the evolution of DM abundance, $Y_\chi(=N_\chi/s a^3)$, for the Yukawa couplings $10^{-1}$, $10^{-3}$, and $10^{-5}$, respectively. Here, the solid and dashed lines demonstrate the \ACCURATE and \APPROXIMATE results, respectively.
Additionally, note that the  \ACCURATE  and  \APPROXIMATE results for Yukawa coupling $10^{-1}$ (green lines) since for such a large coupling, mediator domination never occurred as pointed out by the green star in Figure~\ref{constraint}. In Figure~\ref{Energy_evo}, we have already shown that although there is a prolonged period of early matter-dominated era for $y=10^{-5}$, for  $y=10^{-3}$, the same occurs for a very short period as the mediator decays totally just after starting to dominate the universe's energy budget. Additionally, there is a visible kink in the $Y_\chi$ evolution at around $a/a_\text{RH} \sim 200$ (shown at the inset of Figure~\ref{abundance}), reflecting the fact that the scaling of energy density for the mediator changes as the relativistic to non-relativistic transition takes place at $T \sim m_X$ (compare, Figure~\ref{Energy_evo}). As can be anticipated, the production would stop at the point of bending if mediator roles are neglected, and provide the same DM that is obtained for the standard case. However, a significant production occurs in the non-relativistic phase of the mediator, where it can induce a matter-dominated phase for a period, depending on the smallness of the coupling, which is represented by the change in slope. Furthermore, DM production in the case of matter domination suffers from a dilution as a result of entropy injection by the mediator before its abundance saturates. The DM masses satisfying the relic density constraint are 0.06 keV (0.06 keV), 0.2 keV (0.63 MeV), and 7.4 keV (6.3 GeV) for the Yukawa couplings $10^{-1}$, $10^{-3}$, $10^{-5}$, respectively, for an \ACCURATE (\APPROXIMATE) calculation.

\begin{figure}[tb!]
	\centering
	\includegraphics[width=4cm, height=1.5cm]{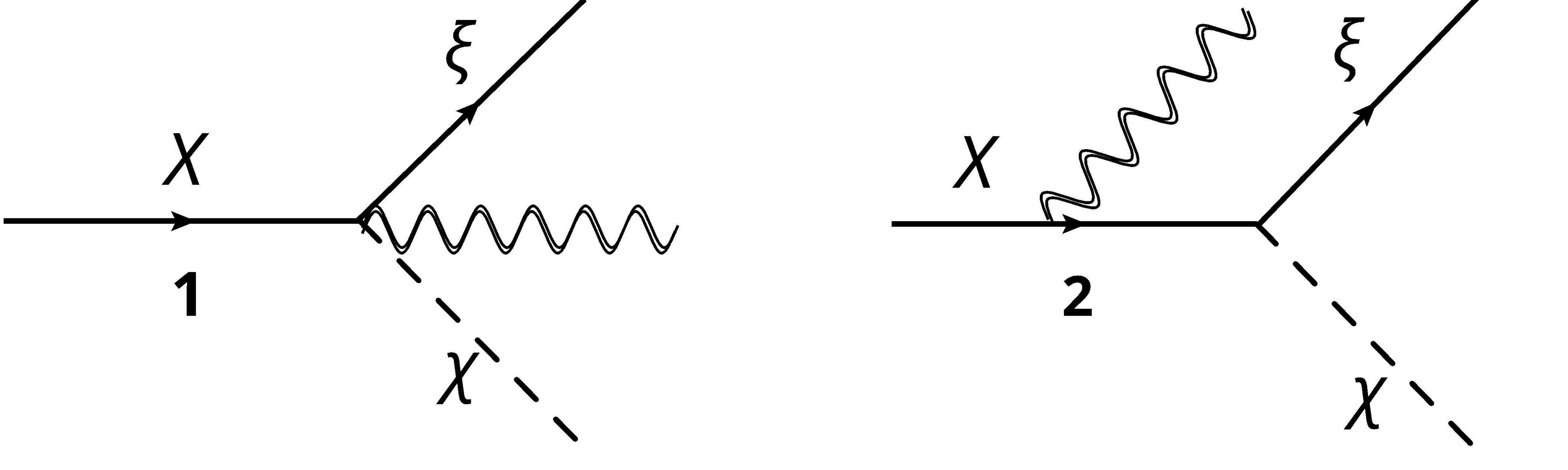}~
	\includegraphics[width=4cm, height=1.5cm]{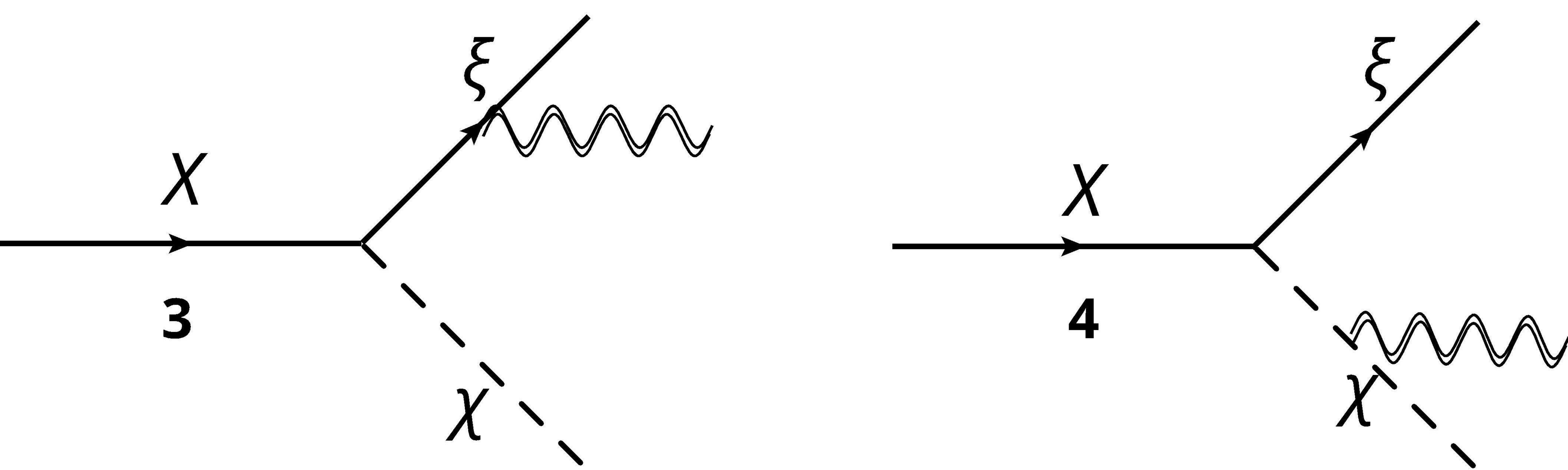} 
	\caption{ Feynman diagrams for graviton production.}
	\label{GW_prod}
\end{figure}

\textbf{\textit{Graviton bremsstrahlung as gravitation wave signature}---} 
The interaction between the canonically normalized graviton $(h_{\mu\nu})$ and the stress-energy tensor $T_i^{\mu\nu}$ for any particle $i$, which denotes SM as well as BSM particles, can be written as~\cite{Choi:1994ax, Holstein:2006bh}
\begin{equation}
	\mathcal{L}_{\text{int}}^{\text{grav}}\supset-\frac{2}{M_P}	h_{\mu\nu}\sum_iT_i^{\mu\nu},
\end{equation}
In particular, the stress-energy tensors for fermion ($i=f$) and scalar ($i=\phi$) can be expressed as
\begin{align}\nonumber
	&T_f^{\mu\nu}=\frac{i}{4}[\bar f\gamma^\mu\partial^\nu f+\bar f\gamma^\nu\partial^\mu f]-\eta^{\mu\nu}\bigg[\frac{i}{2}\bar f\gamma^\sigma\partial_\sigma f -m_f \bar f f\bigg],\\
	& T_\phi^{\mu\nu}=\partial^\mu\phi\partial^\nu\phi-\eta^{\mu\nu}\bigg[\frac{1}{2}\partial^\sigma\phi\partial_\sigma\phi-V(\phi)\bigg],
\end{align}
where $V(\phi)$ corresponds to the potential of the scalar field $\phi$. The graviton production occurs via the graviton \textit{bremsstrahlung} through three-body decay of the heavy particle $X$ without affecting the DM production via the 2-body decay of the same. Figure~\ref{GW_prod} depicts the corresponding Feynman diagrams. In the limit $m_\chi, m_\xi\ll m_X$, the differential decay width for the $1\rightarrow 3$ process with graviton \textit{bremsstrahlung} can be expressed as~\cite{Murayama:2025thw}
\begin{align}
	\frac{d\Gamma^{1\rightarrow 3}}{dE_{\text{gw}}}=\frac{y^2}{512\pi^3}\frac{m_X^2}{M_P^2}\mathcal{G}(x),
	\label{diff_decay_width}
\end{align}	
Here, $\mathcal{G}(x)=(x-2)^2(1-x)^2/x$ with $x=2E_{\text{gw}}/m_X$ and $E_{\text{gw}}(=2\pi f)$, the energy (frequency) of the graviton, spans over an range $0\le E_{\text{gw}}\le m_X/2$ (i.e. $0\le x\le 1$).

From Figure~\ref{GW_prod}, it is clear that the three-body decay of $X$ injects energy to DM and SM sector in addition to GWs generation. We need to separate out the contribution of dumped energy to graviton to study the the energy density of graviton. Now, the 3-body decay rate of $X$ can be decomposed as~\cite{Barman:2023ymn, Kanemura:2023pnv}
\begin{equation}
	\Gamma^{1\rightarrow 3}=\int\frac{d\Gamma^{1\rightarrow 3}}{dE_{\text{gw}}}\frac{m_X-E_{\text{gw}}}{m_X}dE_{\text{gw}}+\int\frac{d\Gamma^{1\rightarrow 3}}{dE_{\text{gw}}}\frac{E_{\text{gw}}}{m_X}dE_{\text{gw}},
\end{equation} 
Here, the second term on the \textit{r.h.s.} of this equation quantifies the energy injected into the graviton from the decay of $X$.
The differential decay rate of the graviton is determined by two parameters, $y$ and $m_X$ (see Eq.~(\ref{diff_decay_width})), which are solely fixed by the DM relic density constraint, as a consequence, the GW spectrum and freeze-in are intrinsically connected. The following Boltzmann equation governs the production as well as evolution of the energy density of GWs in the form of graviton radiation:
\begin{equation}
	\frac{d\rho_{\text{gw}}}{dt}+4H \rho_{\text{gw}}=\bigg[\int\frac{d\Gamma^{1\rightarrow 3}}{dE_{\text{gw}}}\frac{E_{\text{gw}}}{M_X}dE_{\text{gw}}\bigg] \rho_{X},
	\label{GW_evol}
\end{equation}
It is important to note that one can obtain the evolution of GWs by solving only the above equation, considering the equilibrium density for $X$ ($\therefore\rho_{X}=\rho_{X}^\text{eq}$). However, to get the accurate result, one needs to solve Eq.~(\ref{GW_evol}), along with Eqs.~(\ref{BEs}) simultaneously. Here, the production of GWs takes place via the three-body decay and saturates at $t_\text{s}$ or $T_s$. Such a saturation temperature $T_s$ corresponds to $T_\text{FI}(\sim m_X)$ for the standard case, i.e., the scenario without matter domination. In contrast, in the case of induced matter domination, the same saturation in GWs production occurs at $T_D(=T_s)$, dictating the temperature when the mediator decays completely. Notably, it is more useful to express the evolution equation for the differential GW energy density (${d\rho}/{dE_{\text{gw}}}$) as various GW detectors target different frequency ranges. Now, one can compute the present (denoted by subscript) GW spectrum by using $\Omega_{\text{gw}}h^2=h^2(1/\rho_0)d\rho/d\ln E_{\text{gw}}^0$. One can obtain the quantity $d\rho/d\ln E_{\text{gw}}$ at today by getting the same at the saturation point by solving Eq.~(\ref{GW_evol}) first and then considering the redshift effect from the saturation time till today.


Additionally, one can also obtain present day GW relic by solving the following integral
\begin{align}
	\Omega_{\text{gw}}h^2=\Omega_{\gamma}^0h^2\frac{y^2}{512\pi^3}\frac{m_X^2}{M_P^2}\int dt\frac{\rho_X(t)}{\rho_\gamma^0}\frac{a^4}{a_0^4}\bigg(\frac{x^2m_X}{2}\bigg)\mathcal{G}(x),
\end{align}
that provides a approximate analytical expression for the GW relic as
\begin{align}
	\Omega_{\text{gw}}h^2\simeq\Omega_{\gamma}^0h^2\frac{y^2}{512\pi^3}\frac{m_X^2}{M_P^2}\frac{\rho_X^s}{\rho_\gamma^0}\frac{a_s^4}{a_0^4}\bigg(\frac{x_s^2m_X}{2}\bigg)\mathcal{G}(x_s)t_s,
\end{align}
Here, the subscript and superscript $s$ denotes the corresponding quantity at $t_s$ when DM production ceases and $E^s_{\text{gw}}=E^0_{\text{gw}}\frac{a_0}{a_s}$. Now, one can obtain the GW spectrum for radiation domination using the fact that $t_s=1/(2H_\text{RD})$ with $H_\text{RD}$ being the Hubble rate for radiation domination and equilibrium energy density of $X$ at $t_s$. In case of matter domination, one can utilize instantaneous decay approximation i.e., $t_s=t_D=1/\Gamma^{1\rightarrow2}$.
Interestingly, the GW spectrum has a peak which occurs when $\mathcal{G}(x)$ is minimum for $x_\text{peak}=0.36$ which set the following peak frequency for the induced matter dominated scenario
\begin{align}
	f_\text{peak}=\frac{x_\text{peak}}{2\pi}\frac{m_X}{2}\frac{a_s}{a_0}=\frac{1.53\times 10^{9}}{y}\sqrt{\frac{m_X}{10^{15}~\text{GeV}}}~\text{Hz}
\end{align}
dependence of which is strikingly different from the one calculated for a radiation-dominated picture ($\sim 1.65\times 10^{10}$ Hz), independent of Yukawa coupling as well as the mediator mass (calculated using $T_{\text{FI}}\simeq m_X/5$).
 The magnitude of the GW spectrum computed at the peak frequency for mediator dominated matter domination is
\begin{align}
	\Omega_\text{gw}h^2(f_\text{peak})=4.46\times 10^{-15} \bigg(\frac{m_X}{10^{15}\text{GeV}}\bigg)^2.
\end{align}
as opposed to one that depends both on the Yukawa coupling and mediator mass ($\Omega_\text{gw}h^2(f_\text{peak})\propto y^2 m_X$) for the scenario without matter domination.
\begin{figure}[tb!]
	\centering
	\includegraphics[width=\linewidth]{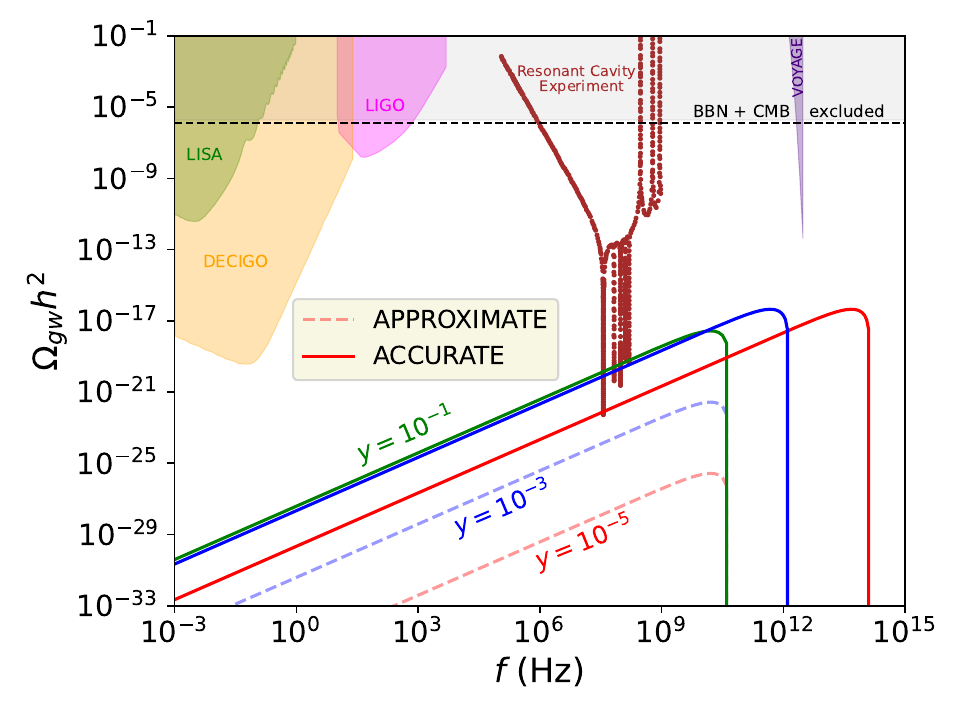}
	\caption{Plot shows GW spectrum.}
	\label{GW_spectrum}
\end{figure}
In Figure~\ref{GW_spectrum}, we have shown the GWs spectrum for the same benchmarks where the green, blue and red plot represent the Yukawa couplings, $10^{-1}, 10^{-3}$ and $10^{-5}$ respectively with the fixed $10^{14}$ GeV mediator mass as chosen in Figure~\ref{abundance}. Here, the solid and dashed line refers to the \APPROXIMATE and \ACCURATE results. Note that these two results merge for the large Yukawa coupling ($y=10^{-1}$) because matter domination will not occur due to the fast decay of the mediator for such a large coupling. 
The future sensitivity ranges of space-based laser interferometer experiments, such as LISA~\cite{amaro2017laser}, DECIGO~\cite{Seto:2001qf}, and LIGO~\cite{KAGRA:2013rdx} operating in the intermediate frequency range ($10^{-6}$-$10^4$ Hz), are indicated by the color shaded regions, respectively. Additionally, the proposed high-frequency experiments, such as the resonant cavity experiment probing higher frequency ranges spanning $10^{4}$-$10^{9}$ Hz~\cite{Herman:2020wao, Herman:2022fau}, Voyage 2050~\cite{He:2023xoh}, are embedded (shown in brown and olive, respectively). 
Furthermore, the energy density generated by GWs acts as radiation before Big Bang Nucleosynthesis (BBN) and contributes to the effective neutrino count ($N_{\text{eff}}$). Consequently, the lower limit imposed by the combined analysis of BBN and Cosmic Microwave Background (CMB) data~~\cite{Yeh:2022heq} on $\Delta N_{\text{eff}}(\le 0.14)$) rules out the indicated grey hatched region in Figure~\ref{GW_spectrum}. 
The GW computation using the approximated approach indicates that only the benchmark with a large Yukawa coupling can be probed by the cavity experiment. In contrast, the accurate calculation shows that the GW spectrum for all the Yukawa couplings associated with the BPs can be tested by the future high-frequency gravitational wave experiments since the peak amplitude as well as the peak frequency shift towards higher values due to induced matter domination for BPs with a relatively smaller Yukawa coupling.

\textbf{\textit{Summary and outlook}--} 
The freeze-in paradigm of dark matter offers one of the most attractive frameworks for its production in the early universe, characterized by its feeble interaction with the thermal bath that is consistent with the null results from diverse dark matter detection experiments. Such studies, in particular those focusing on freeze-in dark matter production from mediator decays, have largely overlooked the cosmological role of the mediator. This letter demonstrates that the energy density of the mediator can dominate the cosmological background in the early universe, leading to an unavoidable matter-dominated era. Such a mediator-driven matter epoch significantly modifies the dynamics of DM production, resulting in a final relic that is several orders of magnitude larger.

On the detection side, new and promising strategies have emerged to test this feebly interacting freeze-in dark matter, including cosmological observations and gravitational wave probes, especially through future high-frequency GW experiments. The mediator-driven matter domination epoch not only alters the production of gravitational waves but also enhances the GW spectrum to a level that can now be made experimentally viable through high-frequency GW experiments.

The impact of this modified dynamics in the universe’s cosmological evolution not only impacts dark matter production or its corresponding GW signature; these effects can also be relevant to broader areas in cosmology. These findings open new avenues for investigating the impact of the mediator on early universe phenomena, including leptogenesis, baryogenesis, and the equilibration of charged lepton temperatures, among other topics.

\textbf{\textit{Acknowledgment}}---SS is supported by NPDF grant PDF/2023/002076 from the Science and Engineering Research Board (SERB), Government of India

\bibliographystyle{apsrev4-2}
\bibliography{ref}

\end{document}